\documentclass[runningheads,a4paper]{llncs}

\usepackage{amsmath}
\usepackage{amssymb}
\usepackage[numbers]{natbib}
\usepackage{graphicx}
\usepackage{booktabs}
\usepackage{listings}
\usepackage{xspace}
\usepackage{multirow}
\usepackage{hyperref}
\usepackage{tikz}
\usetikzlibrary{arrows,shapes,positioning,fit,backgrounds,arrows.meta}
\usepackage{pgfplots}
\pgfplotsset{compat=1.18}
\usepackage[ruled,vlined,linesnumbered]{algorithm2e}

\newcommand{\pyflow}{\textsc{PyFlow}\xspace}

\lstdefinestyle{python}{
  language=Python,
  basicstyle=\footnotesize\ttfamily,
  numbers=left,
  numberstyle=\tiny,
  frame=single,
  columns=fullflexible,
  breaklines=true,
  tabsize=2,
  mathescape=true,
  literate={⊥}{{$\bot$}}1 {∅}{{$\emptyset$}}1,
}

\lstdefinestyle{pseudocode}{
  basicstyle=\footnotesize\ttfamily,
  numbers=left,
  numberstyle=\tiny,
  frame=single,
  columns=fullflexible,
  mathescape=true,
}

\begin{document}

% === Title ===
\title{PyFlow: An Inter-procedural Static
	Analysis Framework for Python}

% === Author block ===
\author{Zinan Gu \and Haoxiang Yan \and Peisen Yao}
\institute{The State Key Laboratory of Blockchain and Data Security, Zhejiang University, Hangzhou, China\\
  \email{\{guzinan1998, yanhx, pyaoaa\}@zju.edu.cn}
}

\maketitle

% === Abstract ===
\begin{abstract}
	Static program analysis infers program properties automatically. Yet precise interprocedural analysis remains challenging, and dynamically typed languages amplify the difficulty. Python is particularly problematic: dynamic dispatch, first-class functions, metaprogramming, pervasive exceptions, and an object model based on descriptors and attribute-driven lookup collectively impede precise reasoning.
	We present \pyflow, a generic IFDS-based static-analysis framework for Python. \pyflow provides a multi-stage intermediate-representation pipeline and a generic IFDS solver parameterized by abstract domains. Analysis developers implement only the dataflow semantics; the framework constructs the supergraph, performs fixed-point iteration, and caches summaries.
	We implement a taint analysis in \pyflow and evaluate it against eight Python SAST tools (DevSkim, Dlint, Bandit, Bearer, CodeQL, Pysa, Semgrep, and Snyk) on the synthetic and real-world benchmarks from a recent ICSE~'26 study. On the synthetic benchmark, \pyflow achieves the best aggregate recall and F1 score among all nine tools. On the real-world benchmark, it attains the highest recall and F1 score while maintaining precision competitive with taint-based engines. We conclude with lessons learned from building IFDS analyses for Python.
\end{abstract}

% === Body ===

\section{Introduction}
\label{sec:intro}

Python is the most widely used programming language today~\cite{tiobe2026},
powering web backends, data pipelines, AI/ML systems, and automation
infrastructure.  The same features that drive Python's adoption---dynamic
dispatch, metaprogramming, first-class functions, runtime
introspection---make it difficult to analyse statically.  Yet the
need for precise, interprocedural static analysis for Python has never been
greater: security vulnerabilities in web frameworks, data-exfiltration paths
through ML pipelines, and incorrect API usage in complex library ecosystems
all demand tooling that understands dataflow across procedure boundaries.

Existing Python static analysis tools address specific tasks.
Bandit~\cite{bandit} scans for security patterns using abstract syntax tree
(AST) matching; mypy~\cite{mypy} and Pyright~\cite{pyright} check type
annotations with limited interprocedural support; Pylint~\cite{pylint} enforces
coding conventions.  None exposes a reusable interprocedural dataflow solver
that is \emph{context-sensitive} and \emph{flow-sensitive}.
Scalpel~\cite{li2022scalpel}, the first general-purpose Python static analysis
framework, provides CFG construction, SSA representation, and call-graph
construction, but does not expose a generic interprocedural dataflow solver.
To obtain precise dataflow information for Python today, a developer must
either implement a solver from scratch or switch to a different language
ecosystem.

For Java and C/C++, by contrast, mature frameworks exist.  FlowDroid~\cite{artz2014flowdroid}
demonstrated the effectiveness of IFDS-based taint analysis for Android
applications.  Tai-e~\cite{tan2023tai} provides a comprehensive Java framework
combining an IFDS solver, pointer analysis, and pass manager.  PhASAR~\cite{schubert2019phasar}
delivers a generic IFDS/IDE solver for C/C++ built on the LLVM infrastructure,
requiring analysis developers only to specify the dataflow problem---the
framework handles supergraph construction, fixed-point iteration, and
summary-edge caching automatically.
So far, such implementations have not been openly available for Python.

Building an IFDS framework for Python presents challenges that do not arise
in statically-typed languages.  Without nominal types, call-graph construction
is fundamentally imprecise: the same variable may refer to objects of
unrelated types at runtime.  Heap-allocated data structures require alias
information for field-sensitive tracking, but Python's object model---attribute
access via descriptors, container indexing, ``duck typing''---complicates
every layer of the analysis.  Furthermore, no sound static analysis is possible
for Python's full feature set~\cite{livshits2015defense}; a framework must
degrade gracefully outside its supported fragment.

We present \pyflow, a generic IFDS interprocedural dataflow
framework for Python.  The frontend and IR pipeline perform source extraction,
dependency resolution, and class-hierarchy construction to produce a unified
multi-level representation shared by all analyses.  An LLVM-inspired pass
manager provides dependency resolution, caching, and invalidation across the
pipeline, and a pointer analysis supplies the heap alias information.
On this infrastructure, the IFDS solver is parameterized over the dataflow
domain, flow functions, and merge operator, enabling reuse across analysis
clients (taint, typestate, and beyond).  Clients are expressed as standard
IFDS flow functions; the solver handles supergraph construction, fixed-point
iteration, summary-edge caching, and context sensitivity automatically.

We evaluate \pyflow's taint analysis against eight widespread Python
SAST tools (DevSkim, Dlint, Bandit, Bearer, CodeQL, Pysa, Semgrep, and
Snyk) on the synthetic and real-world benchmarks released by the recent
ICSE~'26 study of Python SAST tools~\cite{liu2026sast},
spanning web frameworks, HTTP clients, CLI tools, and data pipelines.
On the synthetic benchmark---$240$ programs ($120$ vulnerable and $120$
patched) covering six CWE categories---\pyflow achieves the best
aggregate precision, recall, and F1-score of all nine tools:
$83.7$\%, $85.8$\%, and $84.8\%$, respectively.  On the real-world benchmark ($108$ CVEs across
$62$ projects), \pyflow attains the highest recall ($48.1$\%) and
F1-score ($57.5$\%) while keeping precision competitive with the
interprocedural taint engines.  A case study on SQL injection detection
in a Flask application demonstrates the practical benefit of
context-sensitive interprocedural taint tracking.

This paper makes the following contributions:
\begin{itemize}
  \item We describe the
    design of the multi-level IR pipeline and generic IFDS solver, which
    together enable interprocedural dataflow analysis for Python.

  \item 
    We share practical lessons from building and debugging IFDS analyses on
    real-world Python programs, covering call-graph imprecision, the cost of
    Python's data model, exceptional control flow, and debugging strategies.

  \item 
    We evaluate \pyflow's IFDS solver on the synthetic and real-world
    benchmarks of the ICSE~'26 study~\cite{liu2026sast} and compare its
    precision and recall against eight existing Python SAST tools.
\end{itemize}

\section{Related Work}
\label{sec:related}

Table~\ref{tab:comparison} characterises existing Python tools across the
dimensions that matter for deep program understanding.  Each tool addresses a
subset of these dimensions.  An
open question is whether interprocedural analysis (IFDS, pointer analysis)
\emph{and} fast pattern matching can be supported in a shared framework.

\smallskip 
\noindent\textbf{Python static analysis frameworks.}
Scalpel~\cite{li2022scalpel} is the nearest work to \pyflow in scope: a
general-purpose Python static analysis framework providing CFG construction,
SSA representation, import-graph analysis, and type inference.  Scalpel does
not expose a generic IFDS solver, pointer analysis, plugin interface, or
optimisation pass manager.  The two frameworks follow competing design
philosophies: Scalpel optimises for single-path analysis simplicity, while
\pyflow targets composable extensibility across multiple precision levels.

\smallskip 
\noindent\textbf{Python bug finding tools.}
Bandit~\cite{bandit} is the de facto standard for Python security analysis,
applying AST pattern matching across 500+ checkers.  Semgrep~\cite{semgrep}
provides multi-language pattern-based matching with limited interprocedural
support.  Neither tool performs flow-sensitive, context-sensitive, or
path-sensitive analysis. 
Mypy~\cite{mypy} and Pyright~\cite{pyright} perform type checking with limited
interprocedural analysis (mainly for type inference).  They are insensitive to
dataflow beyond types and do not support security-oriented analyses such as
taint tracking.  Pylint~\cite{pylint} enforces coding conventions via AST
pattern matching.  \pyflow consumes type annotation information when available
but does not replace type checking; its contribution is the analysis
depth---IFDS, alias, context sensitivity---that these tools lack.
%  \pyflow's AST scanner provides a Bandit-compatible fast path for lightweight scanning, while its IFDS and CPG engines support the deep analyses that pattern-based tools cannot express.

\smallskip 
\noindent\textbf{Static analysis frameworks for other languages.}
Several frameworks for Java and C/C++ share individual mechanisms with \pyflow.
IFDS-based frameworks are the closest in analysis approach.  FlowDroid~\cite{artz2014flowdroid}
demonstrated the effectiveness of IFDS-based taint analysis for Android, and
PhASAR~\cite{schubert2019phasar} provides a generic IFDS/IDE solver for C/C++.
Tai-e~\cite{tan2023tai} combines an IFDS solver with a pass manager and pointer
analysis in a single Java framework, systematically selecting design choices
from Soot, WALA, Doop, and SpotBugs.  \pyflow shares the IFDS mechanism with
this group but targets Python, where alias-analysis precision creates additional
challenges for a composable design.  Soot~\cite{vallee1999soot} pioneered the
pass-manager architecture that \pyflow adapts for Python.  LLVM~\cite{llvm}
provides the pass manager and multi-level IR that inspired \pyflow's design.

\begin{table}[t]
\centering
\caption{Design-dimension comparison of Python static analysis tools.
  $\bullet$ = supported/configurable,
  $\circ$ = partial/limited,
  $-$ = not available.}
\label{tab:comparison}
\small
\setlength{\tabcolsep}{4pt}
\begin{tabular}{@{}lccccc@{}}
\toprule
\emph{Tool} & \emph{Inter-} & \emph{Context-} & \emph{Flow-} &
\emph{Generic} & \emph{Pointer} \\
& \emph{procedural} & \emph{sensitive} & \emph{sensitive} &
\emph{IFDS} & \emph{analysis} \\
\midrule
Bandit        & $-$ & $-$ & $-$ & $-$ & $-$ \\
Semgrep       & $\circ$ & $-$ & $-$ & $-$ & $-$ \\
Mypy          & $\circ$ & $\circ$ & $-$ & $-$ & $-$ \\
Pyright       & $\circ$ & $\circ$ & $-$ & $-$ & $-$ \\
Pylint        & $-$ & $-$ & $-$ & $-$ & $-$ \\
Scalpel       & $\bullet$ & $\circ$ & $\circ$ & $-$ & $-$ \\
\midrule
\pyflow        & $\bullet$ & $\bullet$ & $\bullet$ & $\bullet$ & $\bullet$ \\
\bottomrule
\end{tabular}
\end{table}

\section{System Architecture}
\label{sec:architecture}

\begin{figure}[t]
\centering
\resizebox{0.9\textwidth}{!}{% ============================================================
% PyFlow Architecture Diagram — TikZ
% Layers: Frontend / Multi-level IR / Analysis Engines
% (equal-width groups, no inter-layer arrows)
% Requires: \usetikzlibrary{arrows, positioning, fit, backgrounds}
% ============================================================
\begin{tikzpicture}[
	box/.style={
		rectangle, draw, rounded corners=2pt,
		minimum width=18mm, minimum height=9mm,
		inner sep=3pt, outer sep=0pt,
		font=\small, align=center
	},
	group/.style={
		rectangle, draw, rounded corners=4pt, thick, dashed, inner sep=8pt
	},
	glabel/.style={font=\small\bfseries, inner sep=2pt},
	arrow/.style={->, >=stealth, thick}
	]
	
	% ---------- Frontend (row 1) ----------
	\node[box, fill=orange!15] (loc)   at (0,0)   {Source\\Locator};
	\node[box, fill=orange!15] (dep)   at (2.6,0) {Dep.\\Resolver};
	\node[box, fill=orange!15] (astfe) at (5.2,0) {AST\\Extractor};
	\node[box, fill=orange!15] (class) at (7.8,0) {Class\\Hierarchy};
	\draw[arrow] (loc) -- (dep);
	\draw[arrow] (dep) -- (astfe);
	\draw[arrow] (astfe) -- (class);
	
	% ---------- Multi-level IR (row 2) ----------
	\node[box, fill=green!15] (irast) at (0,-2.5)   {AST};
	\node[box, fill=green!15] (ircfg) at (3.9,-2.5) {CFG/SSA};
	\node[box, fill=green!15] (irdep) at (7.8,-2.5) {PDG/CPG};
	\draw[arrow] (irast) -- (ircfg);
	\draw[arrow] (ircfg) -- (irdep);
	
	% ---------- Analysis Engines (row 3) ----------
	\node[box, fill=red!15, minimum width=44mm] (ifds) at (1.3,-5) {IFDS Dataflow Analysis};
	\node[box, fill=red!15, minimum width=44mm] (pa)   at (6.5,-5) {Pointer Analysis};
	
	% ---------- Group boxes (drawn behind nodes) ----------
	\begin{scope}[on background layer]
		\node[group, fill=orange!5, fit=(loc)(dep)(astfe)(class)] (gfe)  {};
		\node[group, fill=green!5,  fit=(irast)(ircfg)(irdep)]    (gir)  {};
		\node[group, fill=red!5,    fit=(ifds)(pa)]               (gana) {};
	\end{scope}
	\node[glabel, anchor=south west, yshift=1pt] at (gfe.north west)  {Frontend};
	\node[glabel, anchor=south west, yshift=1pt] at (gir.north west)  {Multi-Level IR};
	\node[glabel, anchor=south west, yshift=1pt] at (gana.north west) {Analysis Engines};
	
\end{tikzpicture}}
\caption{\pyflow's layered architecture.}
\label{fig:architecture}
\end{figure}
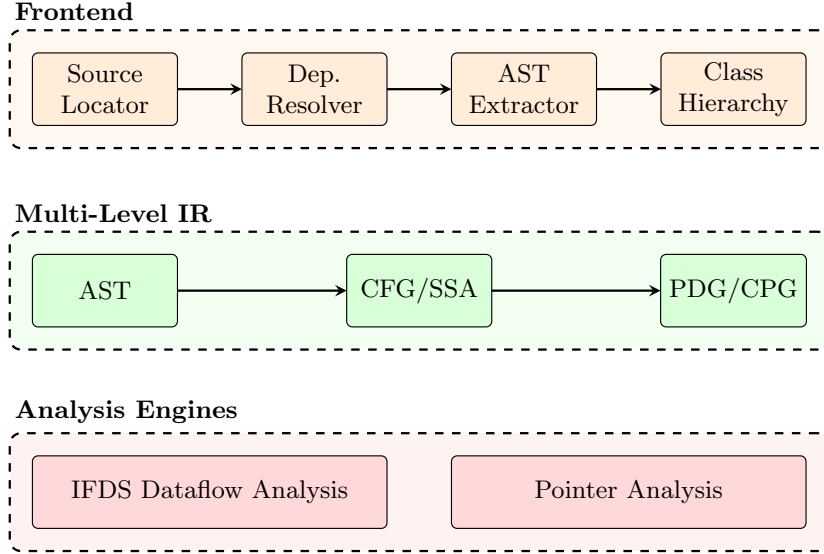

\pyflow's architecture is structured as a layered pipeline.  The frontend
transforms source code into a multi-level intermediate representation;
the optimisation pass manager orchestrates transformations across the
pipeline; and the analysis engines (pointer analysis, IFDS solver,
security clients) consume the shared IR through their respective entry
points.  Figure~\ref{fig:architecture} shows the layered architecture.

\subsection{Frontend and Intermediate Representation}
\label{sec:frontend}

The frontend transforms Python source code into \pyflow's intermediate
representation through five stages.

\smallskip \noindent \textbf{Design rationale.}
The central question for IR design is: \emph{how much analysis should the IR
do?}  Three alternatives exist:

\begin{itemize}
\item \textbf{Raw AST (Bandit, Semgrep).}  Fast to construct and source-faithful
  but provides no control flow, dataflow, or call resolution.
\item \textbf{Linearized IR.}  Efficient for dataflow analysis with explicit
  three-address code and control flow, but loses source structure (loop headers,
  exception regions) and requires non-trivial lowering.
\item \textbf{Multi-level IR.}  Builds a representation chain (AST $\rightarrow$
  CFG $\rightarrow$ SSA $\rightarrow$ dependence graphs) where each level adds
  information and engines choose their entry point.  More expensive up-front,
  but lets precision-flexible engines share construction cost.
\end{itemize}

\pyflow chooses multi-level IR because the cost of constructing all levels
is paid once, and each engine selects its entry point.  An analysis that needs
only raw syntax can consume the AST directly; an IFDS solver that needs the
supergraph can consume the SSA CFG; an analysis that needs the dependence
graph can consume the PDG or CPG.

\smallskip \noindent \textbf{Construction pipeline.}
First, \emph{source location and dependency resolution}: given an entry point
(file or directory), the source locator discovers all relevant Python modules,
and the dependency resolver handles absolute, relative, and namespace imports to
produce a module dependency graph.
Second, \emph{AST extraction}: each module is parsed into a Python AST, which
is then converted into \pyflow's type-annotated IR AST with resolved name
bindings.
Third, \emph{class hierarchy construction}: the class hierarchy analysis
resolves inheritance relationships, metaclass instantiations, and abstract base
class registrations from the IR AST.
Fourth, \emph{control-flow graph generation}: for each function, \pyflow
constructs a CFG with basic blocks, control-flow edges (labeled with
conditions), and exception edges (distinguishing explicit from implicit
exceptions); the CFG is then transformed into SSA form with $\phi$-functions.
Fifth, \emph{dependence graph construction}: from the SSA CFG, \pyflow builds
control dependence graphs (CDG), data dependence graphs (DDG), program
dependence graphs (PDG), and code property graphs (CPG) on demand.

Each stage produces the artifact consumed by the next: the dependency graph
feeds AST extraction, which produces typed ASTs for class hierarchy analysis;
the resolved type information refines CFG construction, and the SSA CFG enables
dependence graph construction.  All IR artifacts are cached and invalidated by
the pass manager when their inputs change.

\subsection{Pass Manager}
\label{sec:passmgr}

The pass manager addresses a question that arises in any framework with
multiple, interacting analyses: \emph{when one analysis depends on the results
of another, how should results be cached, invalidated, and scheduled?}
\pyflow follows LLVM's model~\cite{llvm}: each pass declares its
dependencies and the program artifacts it produces or consumes.  The manager
resolves the pass dependency graph via topological sort, caches pass results,
and invalidates them when dependencies change.
The default pipeline follows a two-pass strategy.  The first pass
establishes baseline results---type information, call targets, escape state---
via alias analysis
passes.

\section{Interprocedural Dataflow Analysis}
\label{sec:ifds}

Dataflow analysis propagates information about a property of
interest---\emph{dataflow facts}---over a program model (typically a
control-flow graph) and captures the effect of each statement on these facts
via \emph{flow functions}~\cite{kam1977monotone, sharir1978two}.  When flow
functions are monotone and distributive with respect to the merge operator,
the analysis can be cast in the Interprocedural Finite Distributive Subset
(IFDS) framework~\cite{repss1995precise} or its generalization, the
Interprocedural Distributive Environments (IDE) framework~\cite{sagiv1996precise}.

IFDS reduces an interprocedural dataflow problem to graph reachability in the
\emph{exploded supergraph} (ESG).  For each interprocedural control-flow node
$n$ and each fact $d \in D$, the ESG contains a node $\langle n, d \rangle$.
Reachability from a distinguished tautological node $\Lambda$ corresponds to
validity: $\langle n, d \rangle$ is reachable iff $d$ holds at $n$.  Unlike
the call-string approach~\cite{sharir1978two}, which requires an a priori
bound $k$, IFDS attains context sensitivity via procedure summaries: once the
solver computes a summary for a given (start-fact, end-fact) pair, it reuses
that summary at all call sites.   
IDE extends IFDS by 
labeling ESG edges with \emph{edge functions} over a secondary value domain $V$, enabling efficient encodings of analyses such as linear constant propagation without inflating the fact domain.  \pyflow currently implements
IFDS; IDE support is deferred to future work.

Despite the widespread adoption of IFDS for Java (e.g., FlowDroid~\cite{artz2014flowdroid}) and C/C++ (PhASAR~\cite{schubert2019phasar}), Python
has lacked an equivalent client-agnostic solver.  Existing Python static analyzers such
as Pysa~\cite{pysa} and CodeQL~\cite{codeql} support taint tracking, but via
special-purpose mechanisms---summary-based fixed-point iteration and ESSA-graph
reachability, respectively---and do not expose a general IFDS/IDE-style
abstraction.  This section presents \pyflow's IFDS implementation, its
adaptation to Python's dynamic semantics, and the analysis clients built on
top of it.

\subsection{Encoding an IFDS Problem}
\label{sec:ifds-encoding}

\pyflow's IFDS solver is parameterized over the supergraph structure, the
dataflow domain~$D$, and four families of flow functions corresponding to the
four edge kinds in the IFDS supergraph, analogous to PhASAR~\cite{schubert2019phasar}
and Heros~\cite{bodden2012inter}:

\smallskip
\noindent\textbf{The \texttt{normal\_flow}.}
Handles intra-procedural propagation.  For an assignment \texttt{x = y}, the
flow function kills facts about~\texttt{x} (strong update) and generates a
new fact binding~\texttt{x} to~\texttt{y}'s abstract value.

\smallskip\noindent\textbf{The \texttt{call\_flow}.}
Maps caller-side facts into the callee's scope at a call site.  The framework
provides a \texttt{MapFactsToCallee} helper that resolves \pyflow's
\texttt{*args} and \texttt{**kwargs} parameter bindings through
Python's argument-matching rules (positional, keyword-only, var-positional,
var-keyword, and default values).

\smallskip\noindent\textbf{\texttt{return\_flow}.}
Maps callee-side facts back to the caller at return sites.  The provided
\texttt{MapFactsToCaller} helper translates callee-locals to the caller's
scope, handling return-value slots and reference/pointer parameters.

\smallskip\noindent\textbf{The \texttt{call\_to\_return\_flow}.}
Propagates facts that are unaffected by a call (stack-local values not passed
as arguments) directly from the call node to the return site, bypassing the
callee's body.

Because flow functions often follow recurring patterns, \pyflow ships a
library of reusable combinators (\S\ref{sec:ifds-combinators}) that analysis
developers can compose rather than writing each function from scratch.

\smallskip 
\noindent \textbf{A concrete example}.
Consider a Flask endpoint that constructs a SQL query via string
interpolation.  The taint analysis instantiates
\texttt{IFDSProblem} with an access-path domain: each fact is a pair
$\langle location,\ access\_path \rangle$ where the access path is a chain
of field or index dereferences.  The initial seed is the tautological fact
$\bot$ (encoded as \texttt{ZeroFact}) at the \texttt{get\_user\_profile}
entry node.  The normal-flow factory for the assignment
\texttt{query = f"SELECT...\{user\_id\}"} generates a taint fact for
\texttt{query} from the tainted \texttt{user\_id}.  The call-flow factory
for \texttt{db.execute(query)} maps the taint fact from the caller's
\texttt{query} variable to the callee's first formal parameter.
Summaries computed for library callees cache the return flow once and replay
it for each call site, keeping the analysis polynomial.

\subsection{Concrete Domain and Combinators}
\label{sec:ifds-combinators}

\pyflow provides a library of reusable flow-function building blocks that
analysis clients compose rather than writing each function from scratch.
Three core combinators---\texttt{IdentityFlow}, \texttt{KillFlow}, and
\texttt{GenFlow}---are the IFDS analogues of the \texttt{Identity},
\texttt{KillAll}, and \texttt{Gen} template classes in
PhASAR~\cite{schubert2019phasar}.  \texttt{IdentityFlow} ensures that the
tautological zero fact~$\bot$ is always propagated (every IFDS problem
requires $\bot \to \bot$ at every node), while non-zero facts pass through
unchanged by default.  \texttt{KillFlow} implements strong update
($\mathit{fact} \to \emptyset$).  \texttt{GenFlow} introduces a new fact
at source assignments.  These combinators are stateless and can be shared
across flow-function factory invocations.

Analysis clients compose these primitives with application-specific logic.
For example, the taint analysis combines \texttt{GenFlow} (to introduce
taint at source calls) with \texttt{IdentityFlow} (to preserve existing
taint across unrelated statements) and custom kill logic (to remove taint
at sanitizer calls).

\subsection{Supergraph Construction}
\label{sec:ifds-supergraph}

The supergraph generalises intraprocedural CFGs by adding call, return, and
call-to-return edges.  \pyflow's \texttt{Supergraph} class
(Figure~\ref{fig:supergraph-api}) provides the query interface that the
solver depends on.  Adapters translate existing IR graphs into this
representation.

\begin{figure}[t]
\centering
\begin{lstlisting}[style=pseudocode]
class Supergraph(Generic[ProcT, NodeT]):
    add_procedure(proc, entry, exits)
    add_normal_edge(src, tgt)        # intra-procedural
    add_call_edge(call_node, callee, return_site=None)
    add_return_site(call_node, return_site)

    is_call_node(node) -> bool
    is_exit_node(node) -> bool
    entry_of(proc) -> NodeT
    ordered_exits_of(proc) -> tuple
    ordered_normal_successors(node) -> tuple
    ordered_callees_of_call_at(node) -> tuple
    ordered_return_sites_of_call_at(node) -> tuple
    ordered_call_to_return_successors(node) -> tuple
\end{lstlisting}
\caption{Supergraph query interface.  The solver accesses the graph only
  through these methods, enabling multiple backends (CFG adapter, synthetic
  graphs, or direct IR construction).}
\label{fig:supergraph-api}
\end{figure}

\smallskip
\noindent\textbf{CFG adapter.}
The \texttt{CFGSupergraphAdapter} lowers PyFlow's existing control-flow
graphs (which use basic-block granularity) into \emph{statement-level} nodes
so that dataflow facts track individual operations rather than coarse blocks.
Each CFG block is decomposed into a sequence of statement nodes, with call
expressions extracted and assigned their own nodes in evaluation order.
The adapter handles:

\begin{itemize}
\item \textbf{Python call semantics:} positional, keyword, keyword-only,
  \texttt{*args}, \texttt{**kwargs}, and default-value bindings are resolved
  by a \texttt{bind\_call\_arguments} function that mirrors Python's
  argument-matching rules.
\item \textbf{Exception edges:} every call node is conservatively treated as a
  potential raise site; handler matching during lowering connects raised
  facts to the appropriate except-block entries.  \texttt{try/except/finally}
  constructs are lowered to explicit fragments with normal, exceptional, and
  abrupt (return/break/continue) exit paths, where finally-blocks are
  duplicated for each exit kind.
\item \textbf{Await/yield boundaries:} suspension points (await, yield,
  yield from) are preserved as explicit control-flow edges so that
  generator and coroutine dataflow is tracked precisely.
\item \textbf{Four-stage call resolution:} callees are resolved through a chain
  of pluggable resolvers---constraint-callgraph edges from the pointer
  analysis, direct-call IR lowering, annotation-based metadata, and
  name-based matching---any of which can produce candidate callees for a
  given call site.
\end{itemize}

% ─────────────────────────────────────────────────────────────────
\subsection{Practical Implementations}
\label{sec:ifds-solver}
%  with bounded call-string context sensitivity (default depth~3)
The IFDS solver follows the classic tabulation algorithm.  The solver
maintains a worklist of path edges $\langle s_p, d_1 \rangle \to
\langle n, d_2 \rangle$ and processes each by dispatching to the appropriate
flow function family based on the node kind.  Context sensitivity follows the
standard IFDS summary mechanism---incoming edges are keyed by (start node,
start fact) and summaries by (start node, exit node, exit fact), providing
unbounded call-string precision for the dataflow solver itself.  \pyflow's
solver additionally tracks a \texttt{CallContext} (a bounded sequence of call
sites) for orthogonal purposes: distinguishing worklist entries during
debugging and tracing, and communicating with the pointer analysis which
operates under a bounded call-string policy.  While the tabulation algorithm
described above captures the core solver
logic, \pyflow's solver implementation adds several practical mechanisms that
are essential for analysing real-world Python programs.

\smallskip 
\noindent\textbf{Budget management.}
Rather than aborting on resource exhaustion, the solver accepts configurable
limits on propagated path edges, runtime, queue size, memory, and other
resources.  When any limit is exceeded, the solver may return a
\emph{partial result}---a fixed point over the explored subgraph with a
status of \texttt{PARTIAL}.  The caller can inspect which nodes were reached
and which limits were hit.  This ``best-effort'' design follows PhASAR's soundy approach~\cite{livshits2015defense}: the analysis degrades gracefully
when confronted with language features or program scales that exceed the
solver's configured capacity.

\smallskip 
\noindent\textbf{Tracing and provenance.}
When tracing is enabled, the solver records a predecessor graph for each
propagated path edge.  After solving, callers can reconstruct an end-to-end
dataflow path by chaining predecessor records backwards from a reached fact
to its seed.  This enables \pyflow's ``explain'' facility---for any reported
finding, the user can request the interprocedural code flow from source to
sink.
Alongside the forward IFDS solver, \pyflow provides a
\texttt{BackwardIFDSSolver} that propagates facts from exit nodes toward
entry nodes, used for analyses such as backward slicing or use-def chain
computation.

\smallskip
\noindent\textbf{Annotation synthesis}.
Many Python programs use C extensions or dynamically-generated code that the
analysis cannot process.  Rather than aborting, \pyflow synthesises
conservative read/write/reference annotations by traversing the AST when the
full CFG pipeline is unavailable.  \pyflow walks each code object's AST, identifies local variable references,
field accesses, and call expressions, and produces annotations that let the
CFG adapter build a supergraph for the analysed subset.  Synthesised
annotations are marked as approximate in the diagnostic output, and the
pipeline degrades gracefully by reporting structured warnings.
We mirror PhASAR's handling of intrinsic and libc
functions, where missing function definitions default to identity flow~\cite{schubert2019phasar}.
% ─────────────────────────────────────────────────────────────────
\subsection{Analysis Clients}
\label{sec:ifds-clients}

The generic solver is instantiated with three analysis clients:
taint, nullness, and typestate.  Each defines its own abstract domain and
flow functions while sharing the supergraph, solver, and alias information.

\begin{table}[t]
\centering
\caption{IFDS analysis clients and their domain definitions.}
\label{tab:ifds-clients}
\begin{tabular}{@{}lll@{}}
\toprule
\emph{Client} & \emph{Domain} $D$ & \emph{Flow-function patterns} \\
\midrule
Taint     & $\langle location,\ access\_path \rangle$ &
  source $\to$ Gen, sink $\to$ check, sanitizer $\to$ Kill \\
Nullness  & $\langle location,\ nullable? \rangle$ &
  nullable return $\to$ Gen, dereference $\to$ check \\
Typestate & $\langle location,\ state \rangle$ &
  open/acquire $\to$ transition, close/release $\to$ transition \\
\bottomrule
\end{tabular}
\end{table}

\smallskip
\noindent\textbf{Taint analysis.}
Facts are access-path pairs rooted at a storage location (local variable,
field, or container element).  Sources introduce taint via \texttt{GenFlow}
at source-function call sites; sinks check for tainted arguments and report
findings; sanitizers are category-aware (e.g., \texttt{user\_input},
\texttt{file}), removing only matching taint categories.  Collections are
modelled explicitly: mutators (\texttt{append}, \texttt{add}) propagate taint
into the container, accessors (\texttt{get}, \texttt{pop}) extract it.
Expression-level facts track taint on intermediate sub-expression results,
handling patterns like \texttt{sink(source().field)} where the taint path
passes through an intermediate expression rather than a named variable.

Alias information from the pointer analysis enriches the
taint domain in two ways.  First, it resolves field accesses on
heap-allocated objects: when a tainted value is stored into
\texttt{obj.field}, the alias analysis determines which abstract object
\texttt{obj} refers to, enabling field-sensitive taint tracking.  Second, it
propagates taint through container operations---\texttt{dict.get(key)} on a
tainted dictionary produces a tainted return value---by resolving the
container's abstract storage slots.

\smallskip
\noindent\textbf{Typestate analysis} tracks resource lifecycle states via finite-state automata.  Built-in
protocols cover \texttt{open/close} (file), \texttt{acquire/release} (lock),
\texttt{connect/close} (socket), and \texttt{begin/commit/rollback}
(transaction).  Custom protocols can be added through the rule-pack
registry.  Each protocol specifies a set of states and transitions (events);
the analysis generates a fact for each resource object and updates its state
at every event call.  A finding is reported when a method is called on a
resource in an unexpected state (e.g., reading from a closed file).  This
client demonstrates the generic-solver advantage: typestate problems are
structurally different from taint (finite automaton vs.\ access-path
propagation), yet both share the same solver and supergraph infrastructure.

\subsection{Guidelines for Analysing Real-World Python Code}
\label{sec:ifds-guidelines}

Developing a static analysis for Python presents challenges that do not arise
in statically-typed languages.  We share our experience from building and
debugging \pyflow's IFDS clients on production code.

\smallskip 
\noindent\textbf{Call-graph imprecision and its impact.}
Unlike Java or C++, Python provides no nominal type system that can resolve
most call targets statically.  \pyflow's four-stage call resolver
(\S\ref{sec:ifds-supergraph}) chains progressively less precise heuristics,
but even the combined result is an over-approximation that may include
spurious callees.  Each spurious callee adds edges to the supergraph and
increases solver work.  In our experience, the constraint-callgraph resolver
produces the tightest call set for
well-typed code, but for programs that use dynamic dispatch heavily (Django's
metaclass-based model system, Flask's decorator-registered routes), the
annotation-based and name-based resolvers contribute most of the callees.
Analysis designers should therefore structure their flow functions to produce
identity summaries quickly for unexpected callee shapes, avoiding expensive
fact propagation through unfamiliar call targets.

\smallskip 
\noindent\textbf{The cost of Python's data model.}
Python's attribute access, container indexing, and descriptor protocol
translate into complex IR operations that a C/C++ analysis never encounters.
A single statement \texttt{obj.attr[key].method(args)} may lower into five or
more IR nodes (getattr, getitem, method lookup, argument packing, call),
each generating dataflow facts.  Our supergraph adapter decomposes such
expressions into per-operation nodes so that the IFDS solver can track facts
at the granularity of individual sub-expressions, but this increases $|N|$
and therefore solver runtime.  In practice, we found that statement-level
lowering adds a factor of 3--5$\times$ to the node count compared with a
basic-block-level representation, but it is necessary to achieve the
expression-level precision that security analyses require (e.g., distinguishing
\texttt{sink(source().field)} from \texttt{sink(unrelated)}).

%\noindent\textbf{Exceptional control flow is pervasive.}
%PhASAR's evaluation notes that C++'s exceptional control flow increases
%solver work because the same fact must be propagated along both normal and
%exceptional paths~\cite{schubert2019phasar}.  Python's exception model is
%significantly more complex: every operation can raise (\texttt{AttributeError},
%\texttt{TypeError}, \texttt{ValueError}), and the supergraph must conservatively
%model every call node as a potential site for a raise.  In \pyflow's CFG adapter,
%nodes whose operations are call expressions are automatically connected to
%exceptional successors unless a call model proves the call cannot raise.
%We found that exception edges account for 30--60\% of the supergraph edges
%in typical Python programs, and that the solver spends a corresponding
%fraction of its time propagating facts along these edges.  The
%\texttt{include\_exceptional\_edges} flag allows users to disable exceptional
%flow modelling for faster (but unsound) analysis when the false-positive rate
%from missing exceptional paths is acceptable.

\smallskip 
\noindent\textbf{Debugging dataflow analyses.}
Debugging an IFDS analysis on Python code is harder than on C/C++ because
the analysis writer must reason simultaneously about Python's dynamic
semantics, the IR lowering, the supergraph structure, and the flow function
logic.  \pyflow's trace mode records a predecessor graph for every propagated
path edge; the \texttt{explain\_fact} and \texttt{explain\_path} methods on
\texttt{IFDSResult} reconstruct the propagation chain for a reached fact.
We found these tools indispensable for diagnosing missing flows (a fact was
not generated where expected) and spurious flows (a fact propagated through
an unexpected callee).

\section{Evaluation}
\label{sec:eval}

This section evaluates \pyflow{} through two research questions:
\begin{description}
\item[RQ1] Does \pyflow{} achieve higher precision and recall than
  alternative approaches on sensitivity-required benchmarks?
\item[RQ2] How does \pyflow{} perform on real-world Python programs
  of varying size, in terms of end-to-end runtime, memory consumption,
  and concrete vulnerability detection?
\end{description}

\smallskip 
\noindent \textbf{Baselines.}
We compare \pyflow against eight widespread static-analysis tools spanning
pattern-matching scanners, heuristic analyzers, taint engines, and query
engines:
\begin{itemize}
\item \textbf{DevSkim}~v1.0.59~\cite{devskim} (1.0k stars): a Microsoft-maintained
  pattern-based scanner that matches insecure coding idioms in a variety of
  languages; its Python rules flag dangerous standard-library and framework
  calls directly.
\item \textbf{Dlint}~v0.16.0~\cite{dlint} (179 stars): a lightweight, plug-in based
  linting and security tool that analyzes Python source for risky patterns
  using fixed syntactic templates.
\item \textbf{Bandit}~v1.8.0~\cite{bandit} (8.2k stars): an open-source analyzer
  that walks the AST of a Python program and applies heuristic checks (its
  ``plugins''); its rule set is publicly maintained on GitHub.
\item \textbf{Bearer}~v1.49.0~\cite{bearer} (2.7k stars): a SAST tool that combines
  code-pattern rules with lightweight heuristics to flag security issues in
  Python code.
\item \textbf{CodeQL}~v2.20.0~\cite{codeql} (9.9k stars): compiles a program into a
  relational database that analysts query in QL to find vulnerabilities; we
  use its public security queries built by Semmle without customization.
\item \textbf{Pysa}~v0.9.23~\cite{pysa} (7.2k stars): the taint analysis component of
  the Pyre type checker, shipping with source/sink models designed for Python
  applications.
\item \textbf{Semgrep}~v1.101.0~\cite{semgrep} (16.1k stars): a rule-driven scanner
  that matches patterns over a parsed tree; although it supports intraprocedural
  taint, most of its available security rules are pattern-based and open source.
\item \textbf{Snyk}~v1.1293.1~\cite{snyk} (5.6k stars): a commercial engine that
  blends pattern matching with taint analysis and proprietary AI; its rule set
  is not disclosed, yet it is widely used for its convenience and free tier.
\end{itemize}

Specifically, DevSkim, Dlint, Bandit, Bearer, and Semgrep represent the pattern- and
heuristic-based category; CodeQL, Pysa, and Snyk represent
production grade interprocedural taint trackers.  Comparing against all eight
isolates whether \pyflow's IFDS approach offers advantages in precision,
recall, or both, and whether a gap exists relative to mature industrial tools.

\smallskip
\noindent \textbf{Benchmark programs.}
\label{sec:eval-bench}
Following the empirical methodology established by the recent ICSE~'26 study
of Python SAST tools~\cite{liu2026sast}, we evaluate \pyflow{} on two
complementary benchmarks that cover six vulnerability categories drawn from
the 2024 CWE Top~25: Command Injection (CWE-77), Deserialization of Untrusted
Data (CWE-502), Code Injection (CWE-94), Cross-site Scripting (CWE-79), Path
Traversal (CWE-22), and SQL Injection (CWE-89).

\smallskip
\noindent {\emph{Synthetic} benchmark.}
The synthetic dataset comprises $240$ programs---$120$ vulnerable programs
($20$ per category, each at least $80$ lines long and generated by
DeepSeek-V3) together with their $120$ hand-constructed fixed counterparts.
Each vulnerable program contains a single known vulnerability whose trigger
location is annotated at the method level, enabling a controlled, directed
evaluation of each tool's precision and recall (Section~\ref{sec:eval-pr}).

\smallskip
\noindent {\emph{Real-world} benchmark.}
The real-world dataset is the first publicly available, manually verified
Python vulnerability benchmark: it contains $108$ unique CVEs collected from
$62$ popular open-source Python projects, each manually cross-checked against
its patch and advisory to determine the precise vulnerability type and
location.  For each CVE we include the vulnerable program and, when available,
its patched version, yielding $206$ programs in total ($108$ vulnerable and
$98$ patched); the remaining $10$ CVEs are unpatched at the time of writing.
Detection is assessed at the method level: an alert matching the ground-truth
vulnerability type and location in the vulnerable version counts as a true
positive, a report on the patched version as a false positive, and silence on
the vulnerable version as a false negative.

For the current IFDS evaluation snapshot, the harness analyzes all $206$
program versions independently, rather than treating the $108$ CVEs as single
inputs.  This distinction is important: a vulnerable version measures recall,
whereas its patched counterpart measures whether the analysis reports a false
positive.  The $10$ CVEs without a patch therefore contribute only a vulnerable
version.

\smallskip
\noindent \textbf{Environment}.
The reported experiments ran on a machine with an Intel Core i7-13700H CPU
(14 cores, 20 threads) and 32\,GB RAM, running Ubuntu 22.04 and Python 3.11.
We used the pointer analysis with its default context policy ($1$-CFA).  The
real-world timing snapshot reported below is one complete run with a
$600$\,s per-program timeout; a timeout is recorded as no detection rather
than as a partial result.

% ─────────────────────────────────────────────────────────────────
\subsection{Synthetic Benchmark: Precision and Recall}
\label{sec:eval-pr}
 We evaluate precision and recall on the directed synthetic
benchmark with ground-truth labels described above
(Section~\ref{sec:eval-bench}).  Each test case is a small, self-contained
Python program with exactly one source--sink pair and a ground-truth label
(\texttt{T} = vulnerable program, \texttt{F} = patched counterpart).  Because
every vulnerable program is paired with a hand-constructed fixed version, a
report on the patched instance is a false positive and a missed report on the
vulnerable instance is a false negative, mirroring the ICSE~'26 study's
method-level assessment.

\smallskip
\noindent \textbf{Precision.}
On the synthetic benchmark, pattern-matching tools achieve the highest
precision: DevSkim and Dlint report no false positives ($100.0$\%) on
patched programs, but at the cost of very low recall.  The heuristic
tools Bandit ($70.6$\%) and Bearer ($72.4$\%) are more balanced.  Among the interprocedural taint engines, CodeQL ($84.5$\%)
and Semgrep ($83.2$\%) keep precision above $80$\%, while Pysa's
precision ($60.0$\%) reflects its sparse report set.  \pyflow achieves
$83.7$\%, competitive with the taint-based tools and, unlike the
pattern matchers, without sacrificing recall.

\smallskip 
\noindent \textbf{Recall.}
Over the full synthetic suite of $240$ programs, \pyflow{} reports
$103$ true positives, $20$ false positives, $17$ false negatives, and
$100$ true negatives.  It detects $85.8$\% of the injected
vulnerabilities---$103$ of the $120$ vulnerable programs---the highest
recall among all tools.  Snyk ($70.5$\%) and Semgrep ($65.8$\%) follow;
the remaining tools fall below $60$\%.  DevSkim ($8.3$\%) and Pysa
($2.5$\%) detect almost nothing, confirming that pure pattern matching
and sparse taint rules cannot cover real code patterns.  \pyflow's
seventeen missed cases arise from library-code flows with incomplete
stubs and dynamically-constructed attribute names (\texttt{setattr}),
which the pointer analysis does not resolve.  Across the full synthetic
benchmark (Table~\ref{tab:microbench-prf}), \pyflow attains the best
aggregate precision, recall, and F1-score: $83.7$\%, $85.8$\%, and $84.8\%$,
respectively.

\begin{table}[t]
\centering
\caption{Aggregate precision, recall, and F1-score over the
  \emph{synthetic} benchmark of the ICSE~'26 study~\cite{liu2026sast}.
  Bold marks the best value in each column.}
\label{tab:microbench-prf}
\small
\setlength{\tabcolsep}{6pt}
\begin{tabular}{@{}lccc@{}}
\toprule
\emph{Tool} & \emph{Precision} & \emph{Recall} & \emph{F1-Score} \\
\midrule
DevSkim  & \textbf{100.0}\% &   8.3\% & 15.4\% \\
Dlint    & \textbf{100.0}\% &  49.2\% & 65.9\% \\
Bandit   &  70.6\% &  40.0\% & 51.1\% \\
Bearer   &  72.4\% &  59.2\% & 65.1\% \\
CodeQL   &  84.5\% &  40.8\% & 55.1\% \\
Pysa     &  60.0\% &   2.5\% &  4.8\% \\
Semgrep  &  83.2\% &  65.8\% & 73.5\% \\
Snyk     &  91.2\% &  70.5\% & 79.5\% \\
\midrule
\pyflow{} & 83.7\% & \textbf{85.8}\% & \textbf{84.8}\% \\
\bottomrule
\end{tabular}
\end{table}

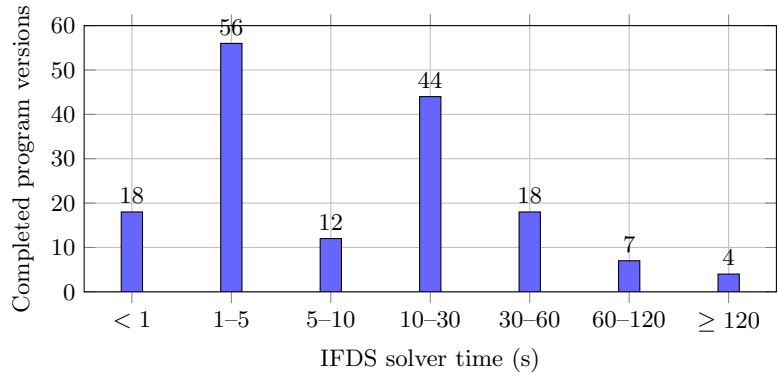
\begin{figure}[t]
	\centering
	\begin{tikzpicture}
\begin{axis}[
  ybar,
  width=0.88\textwidth,
  height=5.1cm,
  bar width=8pt,
  ylabel={Completed program versions},
  xlabel={IFDS solver time (s)},
  symbolic x coords={b1,b2,b3,b4,b5,b6,b7},
  xtick=data,
  xticklabels={$<1$, $1$--$5$, $5$--$10$, $10$--$30$, $30$--$60$, $60$--$120$, $\geq120$},
  xticklabel style={font=\footnotesize},
  ymin=0,
  ymax=60,
  ytick={0,10,20,30,40,50,60},
  enlarge x limits=0.08,
  nodes near coords,
  every node near coord/.append style={font=\footnotesize},
  grid=major,
]
\addplot[fill=blue!60] coordinates {
  (b1,18) (b2,56) (b3,12) (b4,44) (b5,18) (b6,7) (b7,4)
};
\end{axis}
\end{tikzpicture}
	\caption{Distribution of IFDS solver time over the $159$ completed real-world
		program versions.  The $45$ timed-out and $2$ failed versions are excluded
		because no completed solver time is available for them.}
	\label{fig:eval-solver-time}
\end{figure}

\begin{table}[t]
	\centering
	\caption{Precision, recall, and F1-score on the real-world benchmark
		($108$ CVEs across $62$ open-source Python projects) of the ICSE~'26
		study~\cite{liu2026sast}.  Bold marks the best value in each column.}
	\label{tab:realworld-real}
	\small
	\setlength{\tabcolsep}{6pt}
	\begin{tabular}{@{}lccc@{}}
		\toprule
		\emph{Tool} & \emph{Precision} & \emph{Recall} & \emph{F1-Score} \\
		\midrule
		DevSkim  & \textbf{82.4}\% &  13.0\% &  22.5\% \\
		Dlint    &  72.5\% &  26.9\% &  39.2\% \\
		Bandit   &  60.0\% &  27.8\% &  38.0\% \\
		Bearer   &  62.3\% &  35.2\% &  45.0\% \\
		CodeQL   &  72.3\% &  31.8\% &  44.2\% \\
		Pysa     &   0.0\% &   0.0\% &   0.0\% \\
		Semgrep  &  65.7\% &  40.7\% &  50.3\% \\
		Snyk     &  80.0\% &  11.9\% &  20.7\% \\
		\midrule
		\pyflow{} & 71.2\% & \textbf{48.1}\% & \textbf{57.5}\% \\
		\bottomrule
	\end{tabular}
\end{table}

\subsection{Performance on Real-World Projects}
\label{sec:eval-scalability}

\noindent \textbf{Detection on real-world code.}
Table~\ref{tab:realworld-real} reports precision, recall, and F1-score on
the real-world benchmark of $108$ CVEs.  As the ICSE~'26 study
found, every tool's effectiveness drops substantially relative to the
synthetic benchmark: no baseline detects more than $40.7$\% of the
injected vulnerabilities.  Pattern-matching tools retain high precision
(DevSkim $82.4$\%, Dlint $72.5$\%, Snyk $80.0$\%) but very low recall
($11.9$--$26.9$\%), while Pysa fails entirely, reporting nothing on the
real-world dataset ($0.0\%$ across all metrics).  \pyflow{} achieves the
highest recall ($48.1$\%) and F1-score ($57.5$\%) of all nine tools,
surpassing the best baseline (Semgrep) by $7.4$~points in recall and
$7.2$~points in F1-score, while keeping precision ($71.2$\%) competitive
with the interprocedural taint engines (CodeQL $72.3\%$, Semgrep
$65.7\%$).  This confirms that the precision gains observed on the
synthetic benchmark translate to real-world vulnerabilities, where
\pyflow{}'s recall advantage is preserved without an excessive false
positive rate.

\smallskip
\noindent \textbf{Scalability.}
We report IFDS solver time, which isolates data-flow solving from harness,
frontend, and reporting overhead.  In the latest complete real-world run with
a $600$\,s per-program timeout, $159$ of the $206$ program versions completed;
$45$ timed out and $2$ failed.  Across the completed versions, the solver
consumed $3{,}312.1$\,s in total (mean $20.8$\,s), with a median of $6.8$\,s,
a 90th percentile of $49.2$\,s, and a maximum of $445.0$\,s.
Figure~\ref{fig:eval-solver-time} shows the resulting long-tailed distribution:
$74$ ($46.5\%$) completed within $5$\,s and $130$ ($81.8\%$) within
$30$\,s, while only four required at least $120$\,s.  Solver time is unavailable
for timed-out analyses because they do not finish the IFDS solve.

% ─────────────────────────────────────────────────────────────────
%\subsection{Threats to Validity}
\label{sec:eval-threats}

The \emph{synthetic} benchmark is generated by an LLM (DeepSeek-V3) and
was constructed following the methodology of the ICSE~'26 study of
Python SAST tools~\cite{liu2026sast}, so it may not reflect the full
distribution of challenges in production code.  All baselines used
default configurations without customisation, which may understate
their best possible precision or recall.

\section{Conclusion}
\label{sec:conclusion}

We presented \pyflow, a generic IFDS framework for context-sensitive,
interprocedural dataflow analysis of Python, including a multi-level IR
pipeline, pointer analysis, and a generic solver.
We evaluated \pyflow's scalability, precision, and recall on real-world
programs and a directed micro-benchmark, comparing against eight existing
Python SAST tools, and presented practical lessons from developing IFDS
analyses for Python.  \pyflow is released as open source to support future
research and tooling.

\section*{Artifact Availability}
The \pyflow{} implementation, including the IFDS engine, analysis models, and
PySASTBench evaluation harness, is publicly available at
\url{https://github.com/ZJU-PL/pyflow}.

% === Bibliography ===
\bibliographystyle{splncs04}
\bibliography{references}

\end{document}